\documentclass[aps,prl,reprint,longbibliography]{revtex4-1}
\usepackage{mathrsfs}
\usepackage{amsmath}
\usepackage{amsfonts}
\usepackage{amssymb}
\usepackage{amsthm}
\usepackage{graphicx}
\usepackage{natbib}
\usepackage{color}
\usepackage{hyperref}
\usepackage{bm}
\usepackage[caption=false]{subfig}
\usepackage{verbatim}

\begin{document}
\title{Spin pumping and shot noise in ferrimagnets: bridging ferro- and antiferromagnets}
\author{Akashdeep Kamra}
\email{akashdeep.kamra@uni-konstanz.de}
\affiliation{Department of Physics, University of Konstanz, D-78457 Konstanz, Germany}
\author{Wolfgang Belzig}
\email{wolfgang.belzig@uni-konstanz.de}
\affiliation{Department of Physics, University of Konstanz, D-78457 Konstanz, Germany}

\begin{abstract}
 A combination of novel technological and fundamental physics prospects has sparked a huge interest in pure spin transport in magnets, starting with ferromagnets and spreading to antiferro- and ferrimagnets. We present a theoretical study of spin transport across a ferrimagnet$|$non-magnetic conductor interface, when a magnetic eigenmode is driven into a coherent state. The obtained spin current expression includes intra- as well as cross-sublattice terms, both of which are essential for a quantitative understanding of spin-pumping. The dc current is found to be sensitive to the asymmetry in interfacial coupling between the two sublattice magnetizations and the mobile electrons, especially for antiferromagnets. We further find that the concomitant shot noise provides a useful tool for probing the quasiparticle spin and interfacial coupling. 
\end{abstract} 




\maketitle


{\it Introduction.} The quest for energy efficient information technology has driven scientists to examine unconventional means of data transmission and processing. Pure spin current transport in magnetic insulators has emerged as one of the most promising candidates\cite{Bauer2012,Kruglyak2010,Chumak2015,Weiler2013}. Heterostructures composed of an insulating magnet and a non-magnetic conductor (N) enable conversion of the magnonic spin current in the former to the electronic in the latter, thereby allowing for their integration with conventional electronics. In conjunction with the technological pull, these low dissipation systems have provided a fertile playground for fundamental physics~\cite{Sonin2010,Takei2017,Kamra2017}.

Commencing the exploration with ferromagnets (Fs), the focus in recent years has been shifting towards antiferromagnets (AFs)~\cite{Gomonay2014,Jungwirth2016,Baltz2016} due to their technological advantages~\cite{Wadley2016}. While a qualitative understanding of some aspects of AFs, such as spin pumping~\cite{Tserkovnyak2002,Cheng2014}, has been borrowed without much change from Fs, the leading order effects in several other phenomena, such as spin transfer torque~\cite{Cheng2014} and magnetization dynamics~\cite{Gomonay2014}, bear major qualitative differences. Thus, several phenomena, already known for Fs, are now being generalized for AFs~\cite{Barker2016B}.

Although ferrimagnets ($\mathcal{F}$s) have been the subject of comparatively fewer works~\cite{Ohnuma2013,Gepraegs2016,Kamra2017}, their high potential is undoubted. The additional complexity of their magnetic structure comes hand in hand with broader possibilities and still newer phenomena. The spin Seebeck effect~\cite{Uchida2010,Xiao2010,Adachi2013} in an $\mathcal{F}$ with magnetic compensation temperature has unveiled rich physics due to the interplay between the opposite spin excitations in the magnet~\cite{Gepraegs2016}. Further studies have asserted an important role of the interfacial coupling between the magnet and the conductor~\cite{Cramer2017}. While yttrium iron garnet is a ferrimagnet and has been the subject of several studies~\cite{Sandweg2010,Heinrich2011,Czeschka2011,Weiler2013,Bauer2012,Chumak2015}, it is often treated as a ferromagnet on the grounds that only the low energy magnons are important~\cite{Barker2016}.

\begin{figure}[t]
\begin{center}
\subfloat[]{\includegraphics[height=50mm]{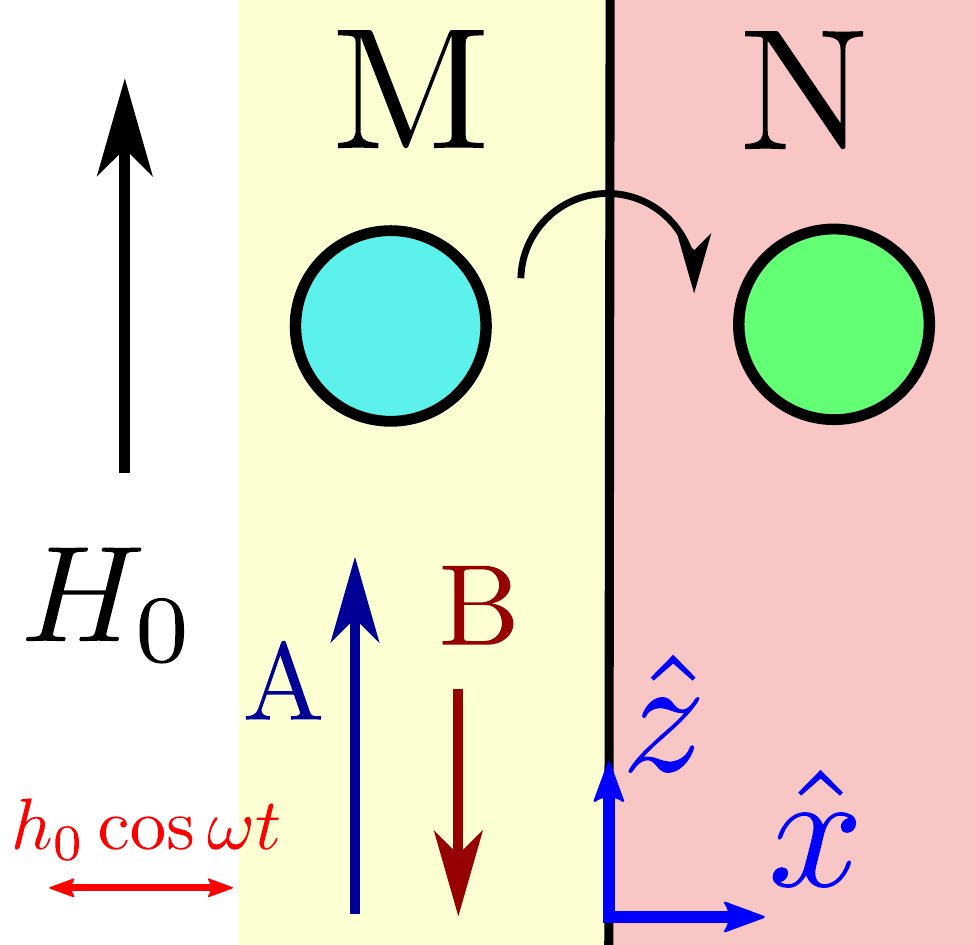}} \qquad
\subfloat[]{\includegraphics[height=50mm]{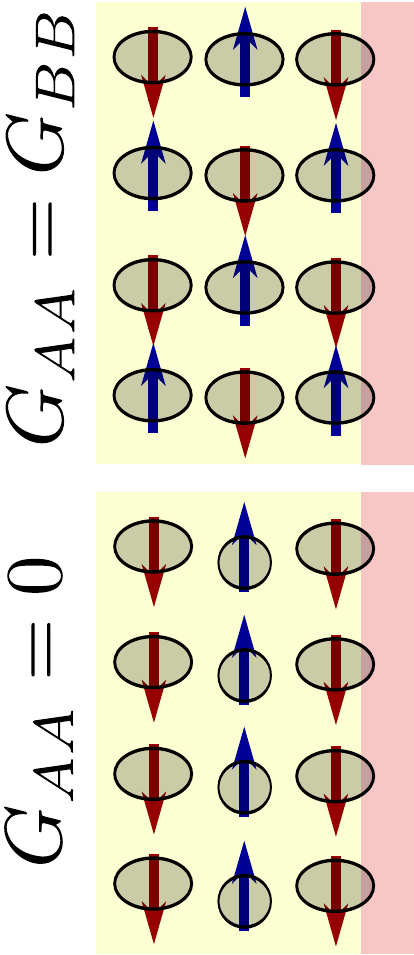}}
\caption{(a) Schematic of the magnet (M)$|$non-magnetic conductor (N) heterostructure under investigation. Equilibrium magnetization for sublattcies A (blue) and B (red) point along $\hat{\pmb{z}}$ and $-\hat{\pmb{z}}$, respectively. An eigenmode in M is driven coherently and injects z-polarized spin current into N. (b) Schematics of possible interface microstructures. Shaded regions around each spin represent the wavefunction cloud of the localized electrons composing the spin. Our model encompasses compensated as well as uncompensated interfaces including lattice disorder.}
\label{fig:bilayer}
\end{center}
\end{figure}

\begin{figure}[t]
   \begin{center}
   \includegraphics[width=85mm]{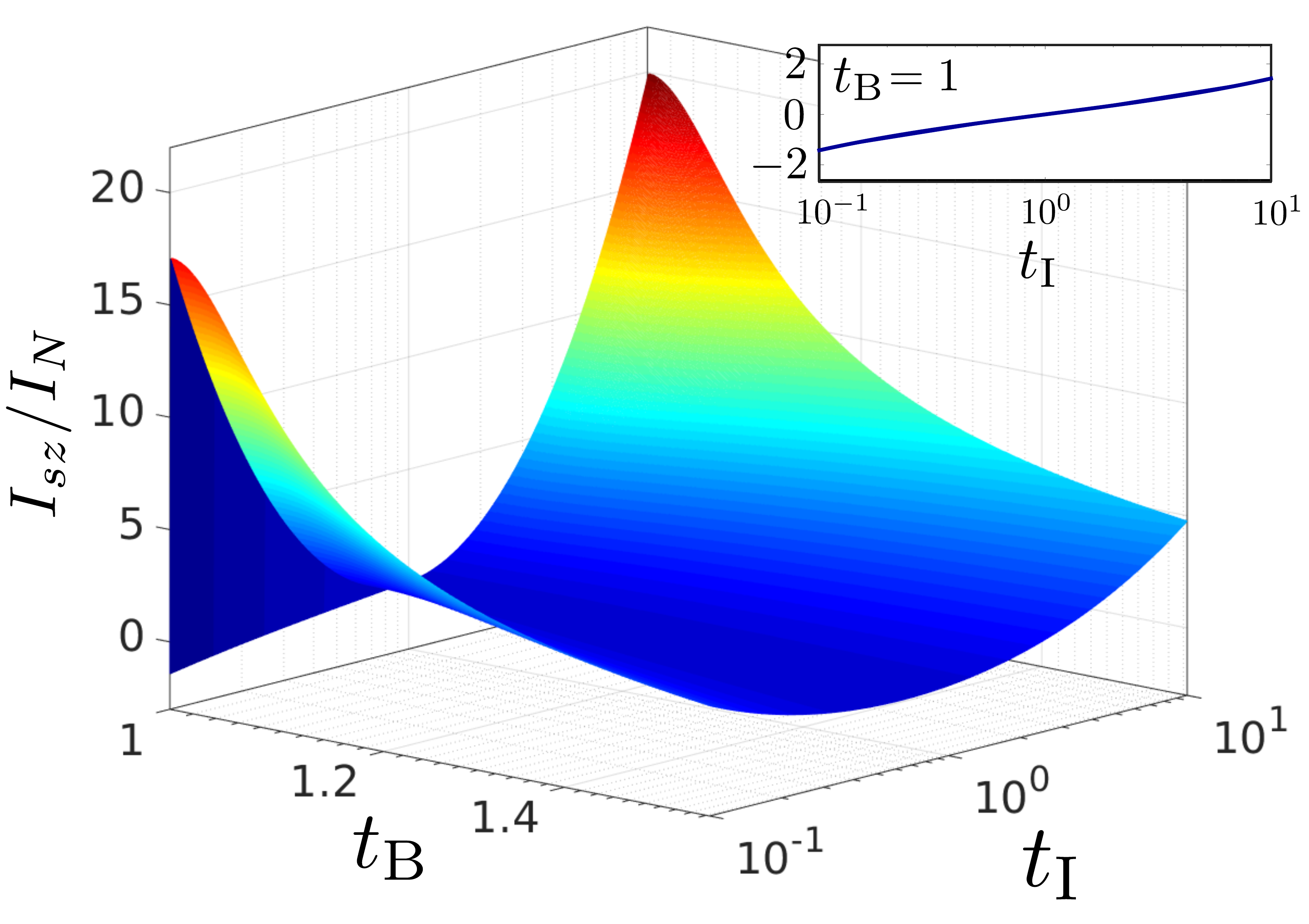}
   \caption{Normalized spin current vs. bulk ($t_B = M_{A0}/M_{B0}$) and interfacial ($t_I = \Gamma_{AA}/\Gamma_{BB}$) asymmetries for lower frequency uniform mode in coherent state. All other bulk parameters are kept constant, no external magnetic field is applied, and $I_N = 2 \hbar |\chi|^2 \omega_{\pmb{q}} \alpha_{AB}$. The spin current for $t_B = 1$ (also depicted in the inset for clarity) is small due to the spin-zero quasiparticles in symmetric AFs, and it abruptly increases with a small bulk symmetry breaking due to quasiparticle transformation into spin $\hbar$ magnons~\cite{Kamra2017}. The different parameter values employed are given in the supplemental material~\cite{SupplMat}.}
   \label{fig:currlmat}
   \end{center}
   \end{figure}

In this Letter, we evaluate the spin pumping current ($I_{sz}$) and the concomitant spin current shot noise [$S(\Omega)$] in a $\mathcal{F}$-N bilayer [Fig. \ref{fig:bilayer}(a)], when one of the $\mathcal{F}$ eigenmodes is driven into a coherent sate. A two-sublattice model with easy-axis anisotropy and collinear ground state is employed. Our model continuously encompasses systems from ferromagnets to antiferromagnets, thereby allowing analytical results for the full range of materials within a unified description. It further allows arbitrary (disordered) interfaces. In addition to the bulk asymmetry, stemming from inequivalent sublattices, we find a crucial role for the interfacial coupling asymmetry (Fig. \ref{fig:currlmat}), consistent with the existing experiments~\cite{Gepraegs2016,Cramer2017} and theoretical proposals~\cite{Bender2017}. Such an asymmetry may occur even in a perfect crystalline interface [Fig. \ref{fig:bilayer}(b)] due to the nature of the termination or the different wavefunction clouds of the electrons constituting the localized spins in the two sublattices. Spin transport in AF-N bilayers is found to be particularly sensitive to the interfacial asymmetry, with spin current nearly vanishing for symmetrical coupling of the two sublattices with N corresponding to the case of a compensated interface (Fig. \ref{fig:currlmat}).

A key result of our work is the following semi-classical expression for the spin current injected into N~\footnote{We emphasize that this expression is restricted to the z-component of spin, and may not be employed for the full spin vector. It has been shown that the microscopic matrix elements corresponding to x and y polarized spin transport are in general different~\cite{Bender2015}. This distinction is often not made in literature.}:
\begin{align}
\frac{e}{\hbar} I_{sz} & = \sum_{i,j = \{A,B\}} G_{ij} (\hat{\pmb{m}}_{i} \times \dot{\hat{\pmb{m}}}_{j})_{z} = \sum_{i,j = \{m,n\}} G_{ij} (\pmb{i} \times \dot{\pmb{j}})_{z}, \label{eq:scurr}
\end{align}
where $\hat{\pmb{m}}_{A (B)}$ is the unit vector along sublattice A (B) magnetization, $\pmb{m} = [ \hat{\pmb{m}}_{A} + \hat{\pmb{m}}_{B} ] / 2$, $\pmb{n} = [ \hat{\pmb{m}}_{A} - \hat{\pmb{m}}_{B} ] / 2$, $G_{mm} = G_{AA} + G_{BB} + 2 G_{AB}$, $G_{nn} = G_{AA} + G_{BB} - 2 G_{AB} $, and $G_{mn} = G_{nm} = G_{AA} - G_{BB}$. Employing $G_{AB} = G_{BA} = \sqrt{G_{AA} G_{BB}}$, which is derived, along with the expressions for $G_{AA}$ and $G_{BB}$, in subsequent discussion below, we further obtain $G_{mm} = (\sqrt{G_{AA}} + \sqrt{G_{BB}})^2$ and $G_{nn} = (\sqrt{G_{AA}} - \sqrt{G_{BB}})^2$. Our result [Eq. (\ref{eq:scurr})] for the injected spin current adds upon the existing understanding of spin pumping via AFs~\cite{Cheng2014} by (i) providing analytic and intuitive expressions for the conductances, (ii) incorporating the cross terms characterized by $G_{AB}$ and $G_{mn}$, (iii) deriving the relation $G_{AB} = \sqrt{G_{AA} G_{BB}}$ based upon a microscopic interfacial exchange coupling model, (iv) accommodating compensated ($G_{AA} = G_{BB}$) as well as uncompensated interfaces, and (v) allowing for interfacial disorder. As detailed in the supplemental material~\cite{SupplMat}, the spin pumping expression given in Ref. \cite{Cheng2014} is recovered from Eq. (\ref{eq:scurr}) by substituting $G_{AB} = G_{BA} = 0$ and $G_{AA} = G_{BB}$, and yields results qualitatively different from what is reported herein~\cite{SupplMat}. This difference in results stems from the assumption made in Ref. \cite{Cheng2014} that $\hat{\pmb{m}}_{A}$ and $\hat{\pmb{m}}_{B}$ are independent variables, which is equivalent to setting $G_{AB} = G_{BA} = 0$ implicitly. $\hat{\pmb{m}}_{A}$ and $\hat{\pmb{m}}_{B}$ are coupled via inter-sublattice exchange and hence cannot be treated as independent when considering system dynamics.

We define the dynamical spin correction factor $S_D$ via the relation $S_D \equiv \lim_{T \to 0} S(0)/ 2 \hbar I_{sz}$, where $T$ is the temperature and $S(0)$ is the low frequency spin current shot noise. When the effect of either the dipolar interaction~\cite{dipnote} or the sublattice coupling on the eigenmode under consideration can be disregarded, $S_D \hbar$ coincides with the spin of the eigenmode. In other words, when a full 4-dimensional (4-D) Bogoliubov transform~\cite{Kamra2017} is required to obtain the relevant eigenmode, $S_D$ is a property of the entire heterostructure and depends upon the bulk as well as the interface. Thus, shot noise offers a useful experimental probe of the interfacial properties as discussed below.


{\it Model.} The model we study consists of a two-sublattice magnet coupled via interfacial exchange interaction to a non-magnetic conductor [Fig. \ref{fig:bilayer}(a)]. We assume $M_{A0} \geq M_{B0}$ with the respective sublattice saturation magnetizations $M_{A0},M_{B0}$. The bulk of the magnet is characterized by a classical free energy density which is then quantized, using the Holstein-Primakoff transformations~\cite{Holstein1940,Kittel1963,Akhiezer1968}, to yield the magnetic contribution to the quantum Hamiltonian $\tilde{\mathcal{H}}_{M}$ in terms of the magnon ladder operators.

We consider Zeeman ($H_{\mathrm{Z}}$), easy-axis anisotropy ($H_{\mathrm{an}}$), exchange ($H_{\mathrm{ex}}$) and dipolar interaction ($H_{\mathrm{dip}}$) (see footnote \cite{dipnote}) in the magnetic free energy density written in terms of the A and B sublattice magnetizations $M_{A}(\pmb{r})$ and $M_{B}(\pmb{r})$. With an applied magnetic field $H_0 \hat{\pmb{z}}$ and $\mu_0$ the permeability of free space, the Zeeman energy density reads $H_{\mathrm{Z}} = - \mu_0 H_0 (M_{Az} + M_{Bz})$. The easy-axis anisotropy is parametrized in terms of the constants $K_{uA},~K_{uB}$ as $H_{\mathrm{an}} = - K_{uA} M_{Az}^2 - K_{uB} M_{Bz}^2$~\cite{Akhiezer1968}. The exchange energy density is expressed in terms of the constants $\mathcal{J}_{A},~\mathcal{J}_{B},~\mathcal{J}_{AB}$ and $\mathcal{J}$~\cite{Akhiezer1968}: $H_{\mathrm{ex}} = \sum_{x_i = x,y,z} [ \mathcal{J}_{ A}  (\partial \pmb{M}_{ A}/ \partial x_{i}) \cdot (\partial \pmb{M}_{ A}/ \partial x_{i}) + \mathcal{J}_{ B} (\partial \pmb{M}_{ B}/ \partial x_{i}) \cdot (\partial \pmb{M}_{ B}/ \partial x_{i}) + \mathcal{J}_{\mathcal{AB}} (\partial \pmb{M}_{ A}/ \partial x_{i}) \cdot (\partial \pmb{M}_{ B}/ \partial x_{i})] + \mathcal{J} \pmb{M}_{ A} \cdot \pmb{M}_{B}$. The dipolar interaction energy density is obtained in terms of the demagnetization field $\pmb{H}_{m}$ that obeys Maxwell's equations in the magnetostatic approximation: $H_{\mathrm{dip}} = - (1/2) \mu_0 \pmb{H}_{m} \cdot (\pmb{M}_{\mathcal{A}} + \pmb{M}_{\mathcal{B}})$~\cite{Akhiezer1968,Kittel1963,Kamra2017}. Quantizing the magnetization fields and employing the Holstein-Primakoff transformation, we obtain the quantum Hamiltonian for the magnet:
\begin{align}\label{eq:HamilM1}
\tilde{\mathcal{H}}_{M}  = & \sum_{\pmb{q}} \left[ \frac{A_{\pmb{q}}}{2} \tilde{a}_{\pmb{q}}^{\dagger} \tilde{a}_{\pmb{q}} + \frac{B_{\pmb{q}}}{2} \tilde{b}_{\pmb{q}}^{\dagger} \tilde{b}_{\pmb{q}} + C_{\pmb{q}} \tilde{a}_{\pmb{q}} \tilde{b}_{-\pmb{q}} + D_{\pmb{q}} \tilde{a}_{\pmb{q}} \tilde{a}_{-\pmb{q}} \right. \nonumber \\
      &  \left.+ E_{\pmb{q}} \tilde{b}_{\pmb{q}} \tilde{b}_{-\pmb{q}} + F_{\pmb{q}} \tilde{a}_{\pmb{q}} \tilde{b}_{\pmb{q}}^{\dagger} \right] + \mathrm{h.c.} \quad ,
\end{align} 
where $\tilde{a}_{\pmb{q}}$ and $\tilde{b}_{\pmb{q}}$ are, respectively, sublattice A and B magnon annihilation operators corresponding to wavevector $\pmb{q}$. Relegating the detailed expressions for the coefficients $A_{\pmb{q}},~B_{\pmb{q}}\cdots$ to the supplemental material~\cite{SupplMat}, we note that $C_{\pmb{q}}$ is dominated by the intersublattice exchange while $D_{\pmb{q}}$, $E_{\pmb{q}}$, $F_{\pmb{q}}$ result entirely from dipolar interaction. The magnetic Hamiltonian is diagonalized via a 4-D Bogoliubov transform to new operators~\cite{Kamra2017} $\tilde{\alpha}_{\pmb{q}} = u_{l\pmb{q}} \tilde{a}_{\pmb{q}} + v_{l\pmb{q}} \tilde{b}_{-\pmb{q}}^\dagger + w_{l\pmb{q}} \tilde{a}_{-\pmb{q}}^\dagger + x_{l\pmb{q}} \tilde{b}_{\pmb{q}} $ and similar for $\tilde{\beta}_{\pmb{q}}$: $\tilde{\mathcal{H}}_{M}  =  \sum_{\pmb{q}} \hbar \omega_{lq} \tilde{\alpha}_{\pmb{q}}^\dagger \tilde{\alpha}_{\pmb{q}} + \hbar \omega_{uq} \tilde{\beta}_{\pmb{q}}^\dagger \tilde{\beta}_{\pmb{q}}$. The subscripts $l$ and $u$ refer to lower and upper modes thus assigning the lower energy to $\tilde{\alpha}$ modes. The diagonal eigenmodes are dressed magnons with spin given by $\hbar (|u_{\pmb{q}}|^2 - |v_{\pmb{q}}|^2 + |w_{\pmb{q}}|^2 - |x_{\pmb{q}}|^2)$~\cite{Kamra2017}. Disregarding dipolar interaction, the eigenmode spin is plus or minus $\hbar$. Incorporating dipolar contribution, the spin magnitude varies between 0 and greater than $\hbar$~\cite{Kamra2017}.

The non-magnetic conductor is modeled as a bath of non-interacting electrons: $\tilde{\mathcal{H}}_N = \sum_{\pmb{k},s = \pm} \hbar \omega_{\pmb{k}} \tilde{c}_{\pmb{k},s}^\dagger \tilde{c}_{\pmb{k},s}$, where $\tilde{c}_{\pmb{k},s}$ is the annihilation operator corresponding to an electron state with spin $s \hbar /2$ along z-direction and orbital wavefunction $\psi_{\pmb{k}}(\pmb{r})$. The conductor is coupled to the two sublattices in the magnet via an interfacial exchange interaction parameterized by $\mathcal{J}_{\mathrm{iA}}$, $\mathcal{J}_{\mathrm{iB}}$:
\begin{align}
\tilde{\mathcal{H}}_{\mathrm{int}} & =  - \frac{1}{\hbar^2} \int_{\mathcal{A}} d^2 \mathcal{\varrho} \sum_{\mathrm{\mathcal{G} = A,B}} \left( \mathcal{J}_{\mathrm{i\mathcal{G}}} \tilde{\pmb{S}}_{\mathrm{\mathcal{G}}}(\pmb{\varrho}) \cdot \tilde{\pmb{S}}_{\mathrm{N}}(\pmb{\varrho})  \right),
\end{align}
where $\pmb{\varrho}$ is interfacial position vector, $\mathcal{A}$ is the interfacial area, $\tilde{\pmb{S}}_{\mathrm{A}}$, $\tilde{\pmb{S}}_{\mathrm{B}} $ and $\tilde{\pmb{S}}_{\mathrm{N}}$ represent spin density operators corresponding to the magnetic sublattices A, B and the conductor, respectively. In terms of the eigenmode ladder operators, the interfacial exchange Hamiltonian reduces to~\footnote{We have retained only the terms which contribute to z-polarized spin transport. The disregarded terms lead to minor shifts in magnon and electron energies, and are important for x and y polarized spin transport~\cite{Bender2015}.}:
\begin{align}
\tilde{\mathcal{H}}_{\mathrm{int}} & =  \hbar \sum_{\pmb{k}_1,\pmb{k}_2,\pmb{q}_1} \left( \tilde{\mathcal{P}}_{\pmb{k}_1\pmb{k}_2\pmb{q}_1} + \tilde{\mathcal{P}}_{\pmb{k}_1\pmb{k}_2\pmb{q}_1}^\dagger  \right), 
\end{align}
where $\tilde{\mathcal{P}}_{\pmb{k}_1\pmb{k}_2\pmb{q}_1} \equiv \tilde{c}_{\pmb{k}_1,+}^\dagger \tilde{c}_{\pmb{k}_2,-} \left( W_{\pmb{k}_1\pmb{k}_2\pmb{q}_1}^A  \tilde{a}_{\pmb{q}_1} + W_{\pmb{k}_1\pmb{k}_2\pmb{q}_1}^B  \tilde{b}_{-\pmb{q}_1}^\dagger \right)$, $\hbar W_{\pmb{k}_1\pmb{k}_2\pmb{q}_1}^{\mathcal{G}} = \mathcal{J}_{\mathrm{i}\mathcal{G}} \sqrt{M_{\mathcal{G}0}/2|\gamma_{\mathcal{G}}|\hbar} \int_{\mathcal{A}} d^2 \pmb{\varrho} [ ~ \psi_{\pmb{k}_1}^* (\pmb{\varrho}) \psi_{\pmb{k}_2}  (\pmb{\varrho}) \phi_{\pmb{q}_1} (\pmb{\varrho}) ]$ with $\gamma_\mathcal{G}$ the typically negative gyromagnetic ratio corresponding to sublattice $\mathcal{G}$ (= A,B), and $\phi_{\pmb{q}_1} (\pmb{r})$ is wavefunction of the magnon eigenmode with wavevector $\pmb{q}_1$. Our goal is to examine the spin~\footnote{In the following discussion, the term `spin' is intended to mean z-component of the spin unless stated otherwise.} current and its noise when one of the magnetic eigenmodes is in a coherent state. We may, for example, achieve the $\alpha_{\pmb{q}}$ mode in a coherent state by including a driving term in the Hamiltonian: $\tilde{\mathcal{H}}_{\mathrm{drive}}  \sim \cos(\omega_{\pmb{q}} t) (\tilde{\alpha}_{\pmb{q}} + \tilde{\alpha}_{\pmb{q}}^\dagger )$~\footnote{A typical method for driving the uniform mode is ferromagnetic resonance. Exciting a non-uniform mode is relatively difficult. Our goal, however, is to understand the nature of individual modes, for which a `theoretical' drive suffices.}. 

The operator corresponding to the z-polarized spin current injected by M into N is obtained from the interfacial contribution to the time derivative of the total electronic spin $(\tilde{\pmb{\mathcal{S}}})$:
\begin{align}\label{eq:spincop}
\tilde{I}_{sz} & = \frac{1}{i \hbar} \left[\tilde{\mathcal{S}}_z, \tilde{H}_{\mathrm{int}} \right]  = \hbar \sum_{\pmb{k}_1,\pmb{k}_2,\pmb{q}_1} \left( - i \tilde{\mathcal{P}}_{\pmb{k}_1\pmb{k}_2\pmb{q}_1} + i \tilde{\mathcal{P}}_{\pmb{k}_1\pmb{k}_2\pmb{q}_1}^\dagger  \right).
\end{align}
The above definition captures the spin pumping contribution to the current injected into N and disregards the effect of interfacial spin-orbit coupling~\cite{Nan2015}. The power spectral density of spin current noise $S(\Omega)$ is given by~\cite{Kamra2016B}: $S(\Omega) =   \int_{-\infty}^{\infty} \mathrm{lim}_{\tau_0 \to \infty}  (1/2\tau_0) \int_{- \tau_0}^{\tau_0}  \langle  \tilde{\delta I}_{sz} (\tau) \tilde{\delta I}_{sz} (\tau - t) +  \tilde{\delta I}_{sz} (\tau - t)  \tilde{\delta I}_{sz} (\tau) \rangle d \tau \ e^{i \Omega t} dt$, where $ \langle ~ \rangle$ denotes the expectation value  and $\tilde{\delta I}_{sz} = \tilde{I}_{sz} - \langle \tilde{I}_{sz} \rangle $ is the spin current fluctuation operator.


\begin{figure}[t]
\begin{center}
\includegraphics[width=85mm]{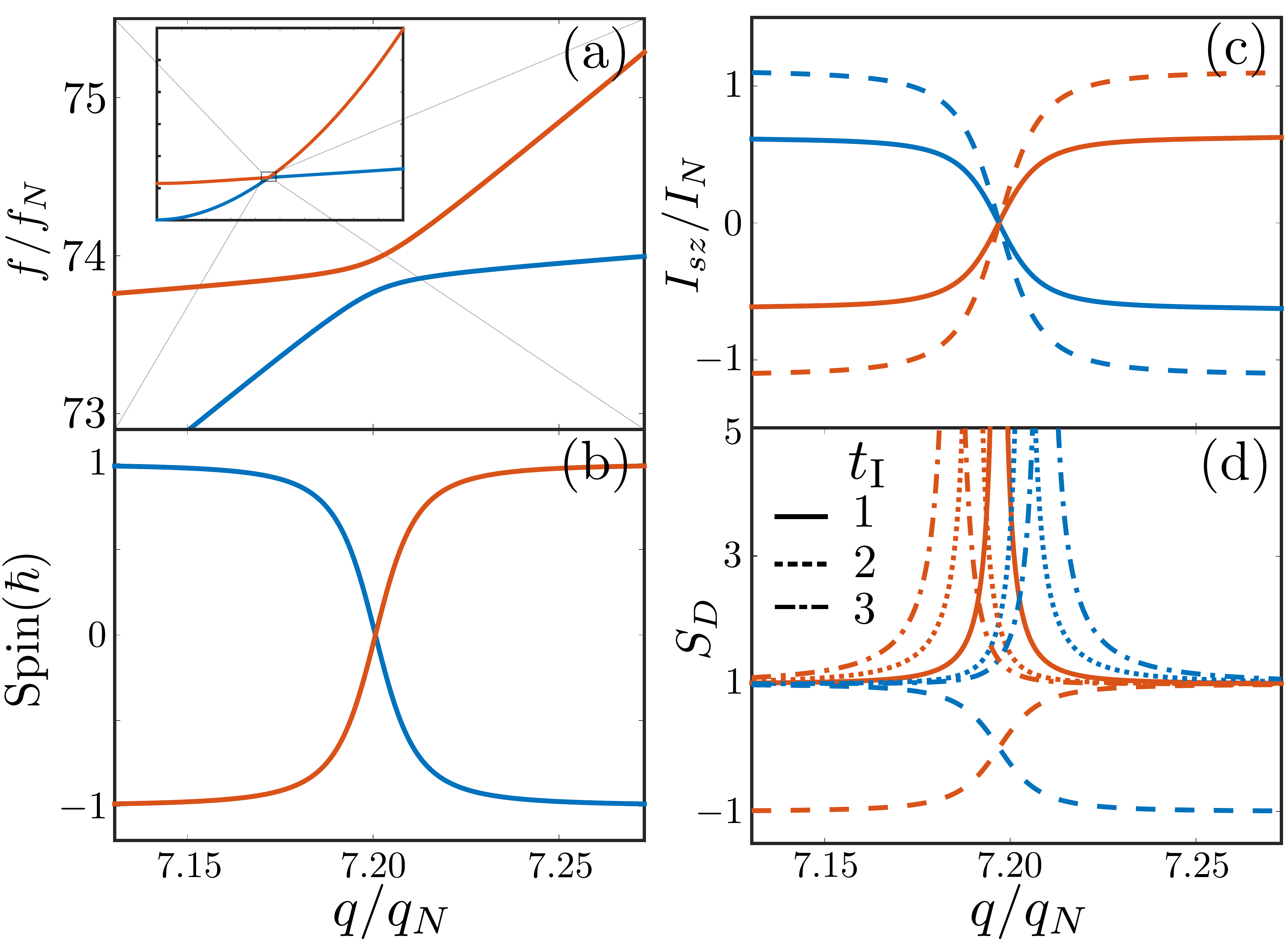}
\caption{(a) Dispersion, (b) quasiparticle spin, (c) spin current injected into N and (d) dynamical spin correction factor vs. wavenumber (along x-direction) around the anti-crossing point in a ferrimagnet. $2 \pi f_N = |\gamma_{\mathcal{A}}| \mu_0 M_{\mathcal{A}0}$ and $f_{l}(q_N) = 2 f_{l}(0)$ define the normalizations $f_N,~q_N$ with $f_{l}(q)$ the lower dispersion band. $I_N = 2 \hbar |\chi|^2 \omega_{\pmb{q}} \alpha_{AB}$ and $t_I \equiv \Gamma_{AA}/\Gamma_{BB} = 1 $, unless stated otherwise. The inset in (a) depicts the full dispersion diagram. Dashed lines in (c) depict the spin current $I_{sz}^\prime$ disregarding the cross-sublattice terms. Dashed lines in (d) depict the quasparticle spin, once again, to help comparison. The parameters employed in the plot are given in the supplemental material~\cite{SupplMat}.}
\label{fig:ferri}
\end{center}
\end{figure}

{\it Results and Discussion.} The spin current $I_{sz}$ in steady state is obtained by evaluating the expectation value of the spin current operator $\tilde{I}_{sz}$ [Eq. (\ref{eq:spincop})] assuming a magnetic mode, e.g. $\alpha_{\pmb{q}}$, in coherent state so that $\tilde{\alpha}_{\pmb{q}}$ may be substituted by a c-number $\chi$~\cite{Kamra2016A}:
\begin{align}\label{eq:scurrquant}
I_{sz}  = &  ~~2 \hbar |\chi|^2 \left[ \Gamma_{AA} \left( |u|^2 - |w|^2 \right) + \Gamma_{BB} \left( |v|^2 - |x|^2 \right) \right. \nonumber \\
     &  \left. - 2 \Gamma_{AB} \Re \left( u^* v - w x^* \right) \right],
\end{align}
where $u,v,w,x$ correspond to the excited eigenmode, $ \Gamma_{ij} = \pi \sum_{\pmb{k}_1,\pmb{k}_2}  W_{\pmb{k}_1\pmb{k}_2\pmb{q}}^{i} \left(W_{\pmb{k}_1\pmb{k}_2\pmb{q}}^{j}\right)^* (n_{\pmb{k}_2} - n_{\pmb{k}_1}) \delta \left(\omega_{\pmb{k}_1} - \omega_{\pmb{k}_2} - \omega_{\pmb{q}} \right)$~\footnote{Note that  $W_{\pmb{k}_1\pmb{k}_2\pmb{q}}^{i} \left(W_{\pmb{k}_1\pmb{k}_2\pmb{q}}^{j}\right)^*$ is real.}, with $i,j$ = $\{A,B \}$, and $n_{\pmb{k}}$ representing the occupancy of the corresponding electron state given by the Fermi-Dirac distribution. Assuming (i) $W_{\pmb{k}_1\pmb{k}_2\pmb{q}}^{\mathcal{G}}$ depends only on the electron chemical potential $\mu$ in N such that it may be substituted by $W_{\mu}^{\mathcal{G}}$, and (ii) the electron density of states around the chemical potential $g(\mu)$ is essentially constant, we obtain the simplified relations: $\Gamma_{ij} = \alpha_{ij} \omega_{\pmb{q}}$. Here, $\alpha_{ij} = \pi \hbar^2 W_{\mu}^{i} (W_{\mu}^{j})^* V_N^2 g^2(\mu)$, with $V_N$ the volume of N. This also entails $\alpha_{AB} = \alpha_{BA} = \sqrt{\alpha_{AA} \alpha_{BB}}$. Since the classical dynamics of a harmonic mode is captured by the system being in a coherent state~\cite{Gerry2004}, the spin current evaluated within our quantum model [Eq. (\ref{eq:scurrquant})] must be identical to the semi-classical expression expected from the spin pumping theory~\cite{Tserkovnyak2002} generalized to a two-sublattice system. As detailed in the supplemental material~\cite{SupplMat}, we evaluate the semiclassical expression given by Eq. (\ref{eq:scurr}) for such a coherent state. The result thus obtained is identical to Eq. (\ref{eq:scurrquant}), provided we identify $G_{ij} = (\alpha_{ij} e/\hbar) \sqrt{M_{i0} M_{j0}/|\gamma_{i}||\gamma_{j}|}$. Since $\alpha_{AB} = \sqrt{\alpha_{AA} \alpha_{BB}}$, we obtain $G_{AB} = G_{BA} = \sqrt{G_{AA} G_{BB}}$~\footnote{The relation $G_{AB} = G_{BA} = \sqrt{G_{AA} G_{BB}}$ holds generally and without making the approximation $W_{\pmb{k}_1\pmb{k}_2\pmb{q}}^{p} \approx W_{\mu}^{p}$.}. These relations along with Eq. (\ref{eq:scurr}) constitute one of the main results of this Letter.  

\begin{figure}[t]
\begin{center}
\includegraphics[width=70mm]{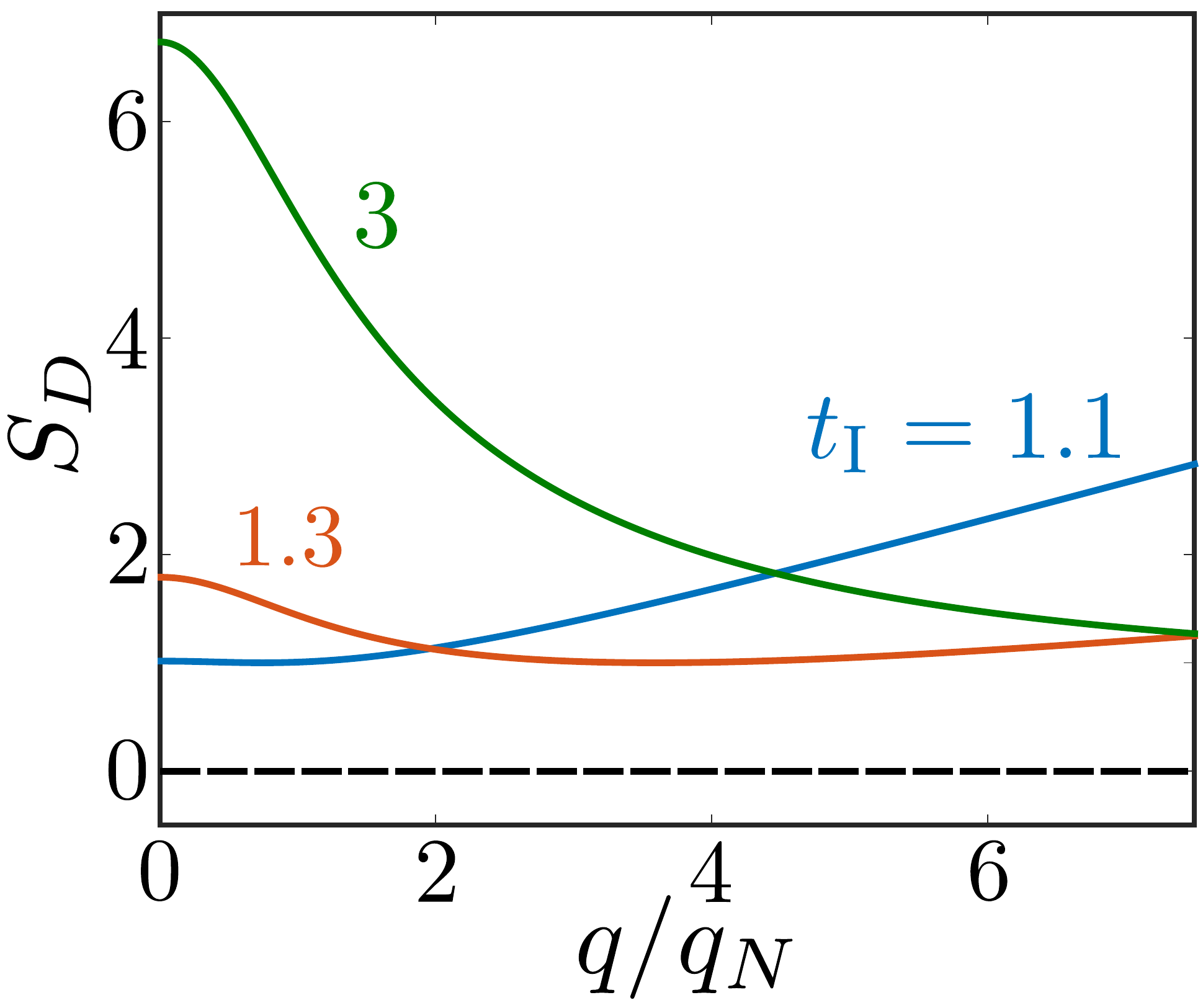}
\caption{Dynamical spin correction factor $S_D$ vs. wavenumber (along x-direction) for a symmetrical AF. Dashed line depicts the zero spin of the magnetic quasiparticles. $f_{l}(q_N) = 2 f_{l}(0)$ defines the normalization $q_N$ with $f_{l}(q)$ the lower dispersion band. The parameters employed in the plot are given in the supplemental material~\cite{SupplMat}.}
\label{fig:spinantiferro}
\end{center}
\end{figure}

In order to gain an understanding of the qualitative physics at play, we examine the injected spin current normalized by $I_N = 2 \hbar |\chi|^2 \omega_{\pmb{q}} \alpha_{AB}$ around the anti-crossing point in the dispersion of a ferrimagnet (Fig. \ref{fig:ferri}) for symmetric interfacial coupling ($\Gamma_{AA} = \Gamma_{BB}$). Due to dipolar interaction~\cite{dipnote}, the dressed magnon spin smoothly changes between plus and minus $\hbar$ resulting in a similar smooth transition in the spin current~\cite{Kamra2017}. Figure \ref{fig:currlmat} depicts the normalized spin current injected by the lower frequency uniform mode ($\pmb{q} = \pmb{0}$) with respect to asymmetries in the bulk $t_B$ ($ = M_{A0}/M_{B0}$) and the interface $t_{I}$ ($= \Gamma_{AA}/\Gamma_{BB}$). For simplicity, we keep all other bulk parameters constant and assume the applied field to vanish. For the case of a perfect AF ($t_B = 1$)~\footnote{The case of an antiferromagnet corresponds to identical parameters for both the sublattices. A compensated ferrimagnet, on the other hand, is represented by identical saturation magnetizations, while the remaining parameters are in general different, for the two sublattices.}, we find a small current with varying $t_I$ that vanishes at $t_I = 1$ (inset in Fig. \ref{fig:currlmat}). The small magnitude of the current is attributed to the dipolar interaction mediated spin-zero magnons in perfect AFs. The spin current has much larger values when $t_B \neq 0$ since the dressed magnons acquire spin $\hbar$ with a small bulk symmetry breaking~\cite{Kamra2017}. The spin current in this case is highly sensitive to $t_I$. This sensitivity is particularly pronounced for AFs, for which the bulk symmetry can also be broken by an applied magnetic field.

The shot noise accompanying the dc spin current injected into N is evaluated for a temperature $T$:
\begin{align}\label{eq:noise}
S(\Omega) & = 2 \hbar |\chi|^2 \left[ \alpha_{AA} \left( |u|^2 + |w|^2 \right) + \alpha_{BB} \left( |v|^2 + |x|^2 \right) \right. \nonumber \\
     &  \left. - 2 \alpha_{AB} \Re \left( u^* v + w x^* \right) \right] [F(\Omega) + F(-\Omega)],
\end{align}
where $F(\Omega) \equiv \hbar (\Omega + \omega_{\pmb{q}} ) \coth(\hbar [\Omega + \omega_{\pmb{q}} ]/[2 k_B T] ) $ with $k_B$ the Boltzmann constant. $F(\Omega) \to \hbar |\Omega + \omega_{\pmb{q}} |$ when $T \to 0$. When the dipolar interaction effect is neglected, i.e. $w,x \to 0$, $\lim_{T \to 0} S(0) \to 2 \hbar I_{sz}$ [Eqs. (\ref{eq:scurrquant}) and (\ref{eq:noise})] such that the dynamical spin correction factor $S_D \to 1$. And when the mode under consideration is not affected by sublattice B, we have $v,x \to 0$ and $S_D \hbar$ approaches the spin of the squeezed-magnon~\cite{Kamra2016A}. In the general case, $S_D~(\geq 1)$ depends upon the magnetic mode, interfacial interaction as well as the eigenmodes in N, and is thus a property of the entire heterostructure. Figure \ref{fig:ferri}(d) depicts $S_D$ for a ferrimagnet around the anti-crossing point in its dispersion. $S_D \approx 1$ away from the anti-crossing, and diverges at some wavenumber which depends upon the interfacial asymmetry $t_I$. This divergence results from a vanishing $I_{sz}$. $S_{D}$ vs. wavenumber for a symmetric AF with varying interfacial asymmetry is depicted in Fig. \ref{fig:spinantiferro}. Thus a combined knowledge of $I_{sz}$ and $S_D$ may allow to probe interfacial asymmetries experimentally~\cite{Kamra2014}. Since deviations of $S_D$ from 1 are necessarily accompanied by quasiparticles with spin different from $\hbar$, it also offers an indirect signature of their formation.

In order to simplify expressions, we have employed the approximation $W_{\pmb{k}_1\pmb{k}_2\pmb{q}}^{\mathcal{G}} \approx W_{\mu}^{\mathcal{G}}$, which is commonly used in the tunneling Hamiltonian description of spin~\cite{Takahashi2010,Zhang2012,Kamra2016A,Kamra2016B} and charge~\cite{Mahan2000} transport. This approximation provides a reasonable description in the limit of strong scattering in N and a disordered interface. The opposite limit of quasi-ballistic transport in N and an ideal AF$|$N interface has been described numerically~\cite{Cheng2014,Takei2014,Bender2017} as well as analytically~\cite{Eirik2017}. Our approximation further disregards the dependence of the spin conductances on $\pmb{q}$~\cite{Kikkawa2015,Ritzmann2015}.

{\it Summary.} We have presented a theoretical discussion of spin transport across a magnet$|$non-magnetic conductor interface when a magnetic eigenmode is driven to a coherent state. Analytical expressions for the dc spin current, including cross terms which were disregarded in Ref. \cite{Cheng2014}, and spin conductances have been obtained. Our theory takes into account the important role of bulk and interfacial sublattice-asymmetries as well as lattice disorder at the interface. The spin current, especially in antiferromagnets, is found to be sensitive to interfacial asymmetry. We have evaluated the spin current shot noise at finite temperatures and shown that it can be employed to gain essential insights into quasi-particle spin and interfacial asymmetry.

{\it Acknowledgments.} We thank Utkarsh Agrawal, So Takei, Scott Bender, Arne Brataas, Ran Cheng, Niklas Rohling, Eirik L{\o}haugen Fj{\ae}rbu, Hannes Maier-Flaig, Hans Huebl, Rudolf Gross, and Sebastian Goennenwein for valuable discussions. We acknowledge financial support from the Alexander von Humboldt Foundation and the DFG through SFB 767 and SPP 1538 SpinCaT. 

{\it Note added in proof.} Recently, Liu and co-workers reported~\cite{Liu2017} a first principles calculation of damping in metallic antiferromagnets. Their conclusions are fully consistent with our work and show the important role of cross-sublattice terms.

\bibliography{spferri_arxiv2}

\begin{thebibliography}{51}
\bibitem [{dip()}]{dipnote}%
  \BibitemOpen
  \href@noop {} {}\bibinfo {note} {Here we use the term ``dipolar interaction''
  to represent any contribution to the magnetic Hamiltonian that results in
  spin non-conserving terms up to the second order in the ladder operators.
  Depending upon the material, these terms may predominantly have different
  physical origin such as magnetocrystalline anisotropy, Dzyaloshinksii-Moriya
  interaction and so on.}\BibitemShut {Stop}%
\end{thebibliography}


\widetext
\clearpage
\setcounter{equation}{0}
\setcounter{figure}{0}
\setcounter{table}{0}
\makeatletter
\renewcommand{\theequation}{S\arabic{equation}}

\begin{center}
\textbf{\large Supplementary material with the manuscript Spin pumping and shot noise in ferrimagnets: bridging ferro- and antiferromagnets by} \\
\vspace{0.3cm}
Akashdeep Kamra and Wolfgang Belzig
\vspace{0.2cm}
\end{center}

\setcounter{page}{1}

\section{Role of cross terms in spin pumping}

\begin{figure}[b]
   \begin{center}
   \includegraphics[width=85mm]{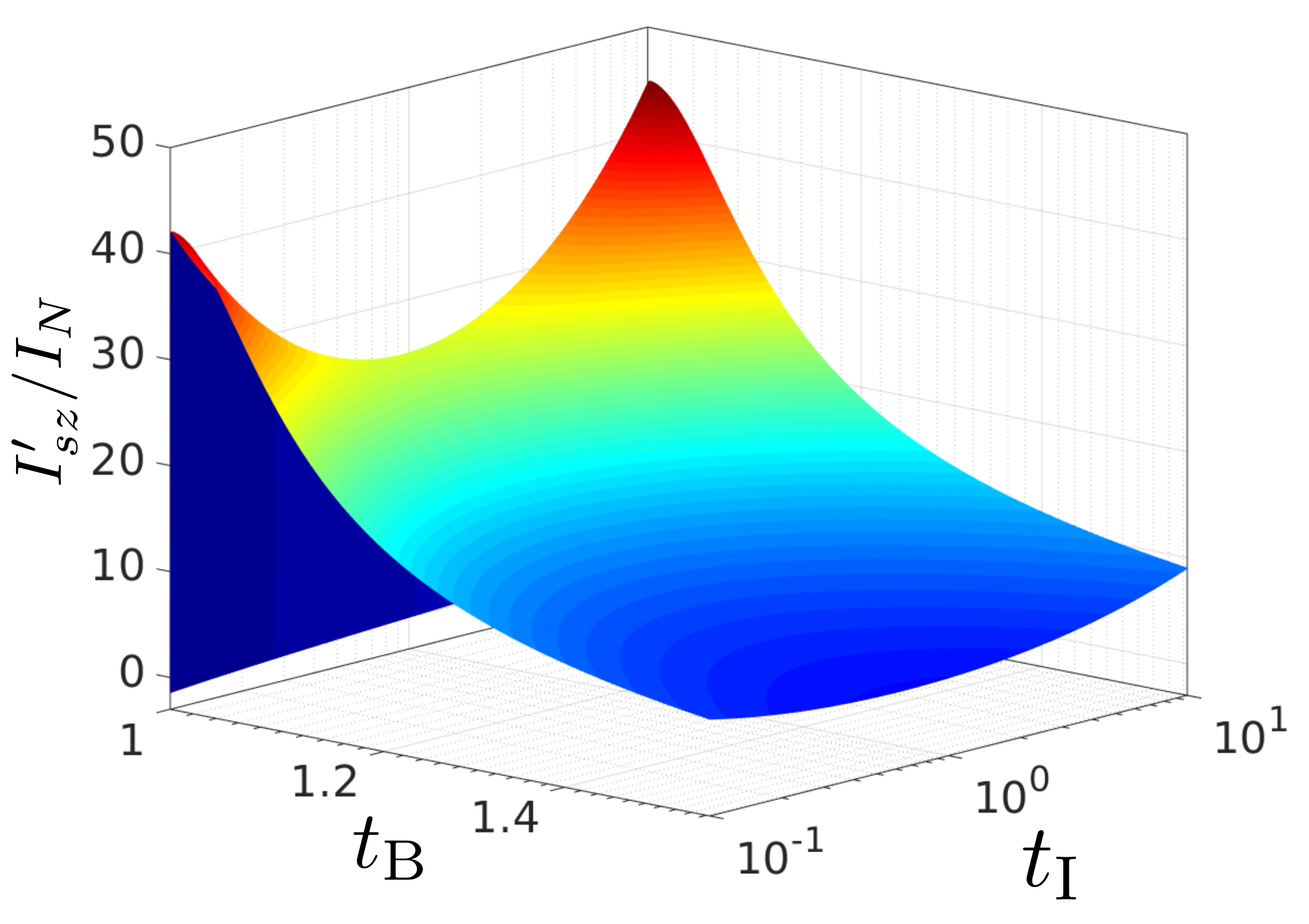}
   \caption{Normalized spin current (disregarding the cross-sublattice terms) vs. bulk ($t_B = M_{A0}/M_{B0}$) and interfacial ($t_I = \Gamma_{AA}/\Gamma_{BB}$) asymmetries for lower frequency uniform mode in coherent state. All other bulk parameters are kept constant, no external magnetic field is applied, and $I_N = 2 \hbar |\chi|^2 \omega_{\pmb{q}} \alpha_{AB}$. The spin current for $t_B = 1$ is small due to the spin-zero quasiparticles in symmetric AFs, and it abruptly increases with a small bulk symmetry breaking due to quasiparticle transformation into spin $\hbar$ magnons~\cite{Kamra2017}.}
   \label{fig:currloldmat}
   \end{center}
   \end{figure}

The semi-classical expression for spin current injected into a conductor ($N$) by an adjacent ferrimagnet ($\mathcal{F}$), when an eigenmode of the latter is driven into a coherent state, is reproduced below (Eq. (1) in the main text).
\begin{align}
\frac{e}{\hbar} I_{sz} & =  G_{AA} (\hat{\pmb{m}}_{A} \times \dot{\hat{\pmb{m}}}_{A})_{z} +  G_{BB} (\hat{\pmb{m}}_{B} \times \dot{\hat{\pmb{m}}}_{B})_{z} +  G_{AB} (\hat{\pmb{m}}_{A} \times \dot{\hat{\pmb{m}}}_{B} + \hat{\pmb{m}}_{B} \times \dot{\hat{\pmb{m}}}_{A})_{z}, \label{eq:scab1}  \\
   & = \frac{e}{\hbar} I_{sz}^\prime + G_{AB} (\hat{\pmb{m}}_{A} \times \dot{\hat{\pmb{m}}}_{B} + \hat{\pmb{m}}_{B} \times \dot{\hat{\pmb{m}}}_{A})_{z},  \label{eq:scab2}
\end{align}
where $\hat{\pmb{m}}_{A (B)}$ is the unit vector along sublattice A (B) magnetization, and we have defined the spin current expression disregarding the cross terms as $I_{sz}^\prime$. Employing $\pmb{m} = [ \hat{\pmb{m}}_{A} + \hat{\pmb{m}}_{B} ] / 2$, and $\pmb{n} = [ \hat{\pmb{m}}_{A} - \hat{\pmb{m}}_{B} ] / 2$, Eq. (\ref{eq:scab1}) can be recast in the following form:
\begin{align}\label{eq:scmn1}
\frac{e}{\hbar} I_{sz}  &  =  G_{mm} (\pmb{m} \times \dot{\pmb{m}})_{z} + G_{nn} (\pmb{n} \times \dot{\pmb{n}})_{z} + G_{mn} (\pmb{m} \times \dot{\pmb{n}} + \pmb{n} \times \dot{\pmb{m}})_{z}, 
\end{align}
where $G_{mm} = G_{AA} + G_{BB} + 2G_{AB}$, $G_{nn} = G_{AA} + G_{BB} - 2G_{AB}$, and $G_{mn} = G_{AA} - G_{BB}$. We note that substituting $G_{AB} = \sqrt{G_{AA} G_{BB}}$, as derived in the main text, yields the expressions for $G_{mm}$, $G_{nn}$ and $G_{mn}$ as specified in the main text. On the other hand, substituting $G_{AB} = 0$ and $G_{AA} = G_{BB}$ leads to an expression ($I_{sz}^\prime$) identical to the one obtained in Ref. \cite{Cheng2014}. To compare the two cases, we plot $I_{sz}^\prime$ vs. bulk and interfacial asymmetries (Fig. \ref{fig:currloldmat}) analogous to the Fig. 2 in the main text. Clear qualitative differences can be seen with $I_{sz}^\prime$ overestimating the injected spin current and underestimating the sensitivity to interfacial asymmetry.

\section{Derivation of the magnetic Hamiltonian}

The classical Hamiltonian for the system is given by the integral of energy density over the entire volume $\mathcal{V}$:
\begin{align}
\mathcal{H}_M & = \int_{\mathcal{V}} d^3 r \ \left(  H_{\mathrm{Z}} + H_{\mathrm{an}} + H_{\mathrm{ex}} + H_{\mathrm{dip}} \right), \\
   & =  \mathcal{H}_{\mathrm{Z}}  + \mathcal{H}_{\mathrm{an}} + \mathcal{H}_{\mathrm{ex}} + \mathcal{H}_{\mathrm{dip}},
\end{align}
with contributions from Zeeman, anisotropy, exchange and dipolar interaction energies, as discussed in the main text. Quantization of Hamiltonian is achieved by replacing the classical variables $\pmb{M}_{A}, \pmb{M}_{ B}$ with the corresponding quantum operators $\tilde{\pmb{M}}_{A}, \tilde{\pmb{M}}_{ B}$. The Holstein-Primakoff (HP) transformation~\cite{Holstein1940,Kittel1963} given by:
\begin{align}
\tilde{M}_{A+}(\pmb{r}) & = \sqrt{2 |\gamma_{A}| \hbar M_{A0}} ~ \tilde{a}(\pmb{r}), \label{eq:hp1} \\ 
\tilde{M}_{ B+}(\pmb{r}) & = \sqrt{2 |\gamma_{ B}| \hbar M_{ B0}} ~ \tilde{b}^{\dagger}(\pmb{r}), \label{eq:hp2} \\ 
\tilde{M}_{A\mathrm{z}}(\pmb{r}) & = M_{A0} - \hbar |\gamma_{A}| \tilde{a}^{\dagger}(\pmb{r}) \tilde{a}(\pmb{r}), \\
\tilde{M}_{ B\mathrm{z}}(\pmb{r}) & = - M_{ B0} + \hbar |\gamma_{ B}| \tilde{b}^{\dagger}(\pmb{r}) \tilde{b}(\pmb{r}), \label{eq:hp4}
\end{align}
expresses the magnetization in terms of the magnonic ladder operators $\tilde{a}(\pmb{r}),~\tilde{b}(\pmb{r})$ corresponding, respectively, to the two sublattices $A,~ B$. In the above transformation, $\tilde{M}_{P+} = \tilde{M}_{P-}^\dagger = \tilde{M}_{Px} + (\gamma_{P}/|\gamma_{P}|) i \tilde{M}_{Py}$, and $\gamma_{P}$, $M_{\mathcal{P}0}$ are the gyromagnetic ratio and the saturation magnetization corresponding to sublattice P. Carrying out the quantization procedure, the magnetic Hamiltonian is obtained:
\begin{align} \label{eq:mham}
\tilde{\mathcal{H}}_M  = & \sum_{\pmb{q}} \left[ \frac{A_{\pmb{q}}}{2} \tilde{a}_{\pmb{q}}^{\dagger} \tilde{a}_{\pmb{q}} + \frac{B_{\pmb{q}}}{2} \tilde{b}_{\pmb{q}}^{\dagger} \tilde{b}_{\pmb{q}} + C_{\pmb{q}} \tilde{a}_{\pmb{q}} \tilde{b}_{-\pmb{q}} + D_{\pmb{q}} \tilde{a}_{\pmb{q}} \tilde{a}_{-\pmb{q}} + E_{\pmb{q}} \tilde{b}_{\pmb{q}} \tilde{b}_{-\pmb{q}} + F_{\pmb{q}} \tilde{a}_{\pmb{q}} \tilde{b}_{\pmb{q}}^{\dagger} \right] + \mathrm{h.c.} \quad ,
\end{align}
where
\begin{align}
\frac{A_{\pmb{q}}}{\hbar} = & \mu_0 H_0 |\gamma_{A}| +  2 K_{\mathrm{u}A} |\gamma_{A}| M_{A0} + 2 \mathcal{J}_{A} q^2 |\gamma_{A}| M_{A0} + \mathcal{J} |\gamma_{B}| M_{B0} \nonumber \\
    & +  \mu_0 |\gamma_{A}|  \left[ N_z (M_{B0} - M_{A0})  + \delta_{\pmb{q},\pmb{0}} \frac{N_x + N_y}{2} M_{A0} + (1 - \delta_{\pmb{q},\pmb{0}}) \frac{\sin^2\left(\theta_{\pmb{q}}\right)}{2} M_{A0} \right] , \\
\frac{B_{\pmb{q}}}{\hbar} = & - \mu_0 H_0 |\gamma_{B}| +  2 K_{\mathrm{u}B} |\gamma_{B}| M_{B0} + 2 \mathcal{J}_{B} q^2 |\gamma_{B}| M_{B0} + \mathcal{J} |\gamma_{A}| M_{A0} \nonumber \\
    & +  \mu_0 |\gamma_{B}|  \left[ N_z (M_{A0} - M_{B0})  + \delta_{\pmb{q},\pmb{0}} \frac{N_x + N_y}{2} M_{B0} + (1 - \delta_{\pmb{q},\pmb{0}}) \frac{\sin^2\left(\theta_{\pmb{q}}\right)}{2} M_{B0} \right], \\
 \frac{C_{\pmb{q}}}{\hbar} = & \sqrt{|\gamma_{A}| M_{A0} |\gamma_{B}| M_{B0}} \left[ \mathcal{J} + \mathcal{J}_{AB} q^2 +  \mu_0 \delta_{\pmb{q},\pmb{0}} \frac{N_x + N_y}{2} + \mu_0 (1 - \delta_{\pmb{q},\pmb{0}}) \frac{\sin^2\left(\theta_{\pmb{q}}\right)}{2}   \right], \\
 \frac{D_{\pmb{q}}}{\hbar} = & \mu_0 |\gamma_{A}| M_{A0} \left[ \delta_{\pmb{q},\pmb{0}} \frac{N_x - N_y}{4} + (1 - \delta_{\pmb{q},\pmb{0}}) \frac{\sin^2\left(\theta_{\pmb{q}}\right)}{4} e^{i 2 \phi_{\pmb{q}}} \right], \\
  \frac{E_{\pmb{q}}}{\hbar} = & \mu_0 |\gamma_{B}| M_{B0} \left[ \delta_{\pmb{q},\pmb{0}} \frac{N_x - N_y}{4} + (1 - \delta_{\pmb{q},\pmb{0}}) \frac{\sin^2\left(\theta_{\pmb{q}}\right)}{4} e^{- i 2 \phi_{\pmb{q}}} \right], \\
  F_{\pmb{q}} = & 2 \sqrt{D_{\pmb{q}} E_{\pmb{q}}^*} .
\end{align}
$N_{x,y,z}$ in the expressions above are the components of the demagnetization tensor in its diagonal form, $\theta_{\pmb{q}},~\phi_{\pmb{q}}$ are respectively the polar and azimuthal angles of $\pmb{q}$, and all remaining symbols have been defined in the main text.

\section{Values of model parameters}
\begin{center}
\begin{tabular}{c | c | c | c | c}
\hline
Parameter & Fig. 2 & Fig. 3 & Fig. 4 & Units \\
\hline
$\mu_0 H_0$ & 0   & 0.05  & 0 & T \\
$N_x$, $N_y$, $N_z$ & 1,0,0  & 1,0,0 & 1,0,0 & Dimensionless \\
$\gamma_{A}$  & 1.8 & 1.8  & 1.8 & $\times 10^{11}$ $\mathrm{s}^{-1} \mathrm{T}^{-1}$  \\
$\gamma_{B}$ & 1.8 & 1.8  & 1.8 & $\times 10^{11}$ $\mathrm{s}^{-1} \mathrm{T}^{-1}$ \\
$M_{A}$ & 5 & 5 & 5 & $\times 10^5$ A/m \\
$M_{B}$  & $M_{A} / t_{B}$  & 2.5 & 5 & $\times 10^5$ A/m \\
$\mathcal{J}_{A}$  &  1 & 5 &  1 & $\times 10^{-23}$ $\mathrm{J}\cdot\mathrm{m} \mathrm{A}^{-2} $   \\
$\mathcal{J}_{B}$ &  1 & 1 &  1 & $\times 10^{-23}$ $\mathrm{J}\cdot\mathrm{m} \mathrm{A}^{-2} $  \\
$\mathcal{J}_{AB}$  &  0.1 & 0.1 &  0.1 & $\times 10^{-23}$ $\mathrm{J}\cdot\mathrm{m} \mathrm{A}^{-2} $  \\
$\mathcal{J}$  & 5  & 1 &  5 & $\times 10^{-4}$ $\mathrm{J} \mathrm{m}^{-1} \mathrm{A}^{-2} $ \\
$K_{uA}$ & 2 & 2  & 2 & $ \times 10^{-7}$ $\mathrm{J} \mathrm{m}^{-1} \mathrm{A}^{-2} $ \\
$K_{uB}$  & 2 & 2  & 2 & $ \times 10^{-7}$ $\mathrm{J} \mathrm{m}^{-1} \mathrm{A}^{-2} $ \\
\hline
\end{tabular}
\end{center}

\section{Semi-classical and quantum expressions for spin current}

A key result of our work is the semi-classical expression [Eq. (\ref{eq:scab1})] for the spin current injected by the ferrimagnet into the conductor in terms of the sublattice magnetizations. This has been derived under the assumption that one magnetic mode is driven into a coherent state. Since a coherent state emulates the classical dynamics of a harmonic oscillator, this semi-classical result should be identical to an analogous expression for spin current obtained within a quasi-classical theory. Here, we demonstrate this equivalence rigorously and identify the spin conductances in terms of the parameters within our microscopic model.

The magnetic Hamiltonian [Eq. (\ref{eq:mham})] can be diagonalized by a four-dimensional Bogoliubov transform~\cite{Kamra2017}:
\begin{align}\label{eq:4dbt}
\begin{pmatrix}
\tilde{\alpha}_{\pmb{\kappa}} \\ 
\tilde{\beta}_{-\pmb{\kappa}}^{\dagger} \\
\tilde{\alpha}_{-\pmb{\kappa}}^{\dagger} \\
\tilde{\beta}_{\pmb{\kappa}}
\end{pmatrix}
 = & 
 \begin{pmatrix}
 u_{1  } & v_{1  } & w_{1  } & x_{1  } \\
 u_{2  } & v_{2  } & w_{2  } & x_{2  } \\
 u_3 & v_3 & w_3 & x_3 \\
 u_4 & v_4 & w_4 & x_4
 \end{pmatrix}
 \begin{pmatrix}
  \tilde{a}_{\pmb{\kappa}} \\ 
  \tilde{b}_{-\pmb{\kappa}}^{\dagger} \\
  \tilde{a}_{-\pmb{\kappa}}^{\dagger} \\
  \tilde{b}_{\pmb{\kappa}}
  \end{pmatrix}  =
  \underline{S} 
  \begin{pmatrix}
 \tilde{a}_{\pmb{\kappa}} \\ 
 \tilde{b}_{-\pmb{\kappa}}^{\dagger} \\
 \tilde{a}_{-\pmb{\kappa}}^{\dagger} \\
 \tilde{b}_{\pmb{\kappa}}
 \end{pmatrix},
\end{align} 
where $\pmb{\kappa}$ denotes the wavevector $\pmb{q}$ running over half space~\cite{Kamra2017,Holstein1940}. The transformation matrix $\underline{S}$ is obtained by imposing the requirement that the Hamiltonian should reduce to:
\begin{align}
\tilde{\mathcal{H}}_M  & =   \sum_{\pmb{\kappa}} [\hbar \omega_{l \pmb{\kappa}} ( \tilde{\alpha}_{\pmb{\kappa}}^{\dagger} \tilde{\alpha}_{\pmb{\kappa}}  + \tilde{\alpha}_{\pmb{-\kappa}}^{\dagger} \tilde{\alpha}_{\pmb{-\kappa}}) +  \hbar \omega_{u \pmb{\kappa}}  (\tilde{\beta}_{\pmb{\kappa}}^{\dagger} \tilde{\beta}_{\pmb{\kappa}} + \tilde{\beta}_{\pmb{-\kappa}}^{\dagger} \tilde{\beta}_{\pmb{-\kappa}})].
\end{align}
Here, we have employed the invariance of the coefficients $A_{\pmb{\kappa}}, B_{\pmb{\kappa}}, \cdots$, appearing in the magnetic Hamiltonian [Eq. (\ref{eq:mham})], under the replacement $\pmb{\kappa} \to -\pmb{\kappa}$. This invariance also leads to the following properties of the transformation matrix $\underline{S}$:
\begin{align}\label{eq:invar}
S_{22} = S_{11}^*, \quad S_{21} = S_{12}^*,
\end{align}
where $S_{ij}$ are the $2 \times 2$ block matrices constituting the $4 \times 4$ matrix $\underline{S}$. Since $\underline{S}$ transforms a set of bosonic operators into a different set of bosonic operators, the corresponding commutation rules impose yet another constraint on the transformation matrix:
\begin{align}\label{eq:boscom}
\underline{S} ~ \underline{Y} ~ \underline{S}^{\dagger} & = \underline{Y}  \implies  \underline{S}^{-1} = \underline{Y} ~ \underline{S}^{\dagger} ~ \underline{Y}^{-1},
\end{align} 
where $\underline{Y} = \sigma_{z} \otimes \sigma_{z}$, with $\sigma_{z}$ the third Pauli matrix.

We consider that the mode $\tilde{\alpha}_{\pmb{q}}$ is in a coherent state so that the operator $\tilde{\alpha}_{\pmb{q}}$ can be replaced by a c-number $\chi$. All other modes are assumed to be in equilibrium. The dynamics of this coherent mode is captured by replacing all quantum operators by their expectation values. Employing Eqs. (\ref{eq:4dbt}), (\ref{eq:invar}) and (\ref{eq:boscom}), we obtain:
\begin{align}\label{eq:abcoh}
\left \langle \tilde{a}_{\pmb{q}} \right \rangle & = u_{1}^* \chi - w_1 \chi^*, \\
\left \langle \tilde{b}_{\pmb{q}} \right \rangle & = x_1^* \chi - v_1 \chi^*. 
\end{align}
The above two equations in conjunction with Eqs. (\ref{eq:hp1}) and (\ref{eq:hp2}) express the expectation values of the magnetization operators. Employing $\chi = |\chi| e^{- i \omega_{\pmb{q}} t}$, we thus evaluate:
\begin{align}
\left ( \left \langle \hat{\pmb{m}}_{A} \right \rangle \times \frac{d}{dt} \left \langle \hat{\pmb{m}}_{A} \right \rangle  \right)_{z} = & 2 \hbar \omega_{q} \frac{|\gamma_{A}|}{M_{A0}} |\chi|^2 (|u_1|^2 - |w_1|^2), \label{eq:res1} \\
\left ( \left \langle \hat{\pmb{m}}_{B} \right \rangle \times \frac{d}{dt} \left \langle \hat{\pmb{m}}_{B} \right \rangle  \right)_{z} = & 2 \hbar \omega_{q} \frac{|\gamma_{B}|}{M_{B0}} |\chi|^2 (|v_1|^2 - |x_1|^2), \\
\left ( \left \langle \hat{\pmb{m}}_{A} \right \rangle \times \frac{d}{dt} \left \langle \hat{\pmb{m}}_{B} \right \rangle  \right)_{z} + \left ( \left \langle \hat{\pmb{m}}_{B} \right \rangle \times \frac{d}{dt} \left \langle \hat{\pmb{m}}_{A} \right \rangle  \right)_{z}= & 2 \hbar \omega_{q} \sqrt{\frac{|\gamma_{A}| |\gamma_{B}|}{M_{A0} M_{B0}}} |\chi|^2 \left[ - 2 \Re \left( u_1^* v_1 - w_1 x_1^* \right) \right]. \label{eq:res2}
\end{align}
The equations (\ref{eq:res1}) - (\ref{eq:res2}) obtained above demonstrate the equivalence between the semi-classical (Eq. (1) in the main text) and the quantum (Eq. (6) in the main text) expressions for the spin pumping current, and allow us to identify the spin conductances in terms of the parameters in the quantum model.

\end{document}